\documentclass[english,aps,prper,reprint,showpacs,titlepage,longbibliography]{revtex4-2}   

\usepackage[T1]{fontenc}	
\usepackage{geometry}
\usepackage{graphicx}
\usepackage{times}
\usepackage{hyperref}  
\hypersetup{colorlinks=true,urlcolor=blue,citecolor=blue,linkcolor=blue}
\usepackage{array}
\usepackage{enumerate}
\usepackage{enumitem}  
\usepackage{amsmath}
\usepackage{amssymb}
\usepackage{tikz}
\usepackage{graphicx}
\usepackage{multirow}
\usepackage{tcolorbox}
\usepackage{ragged2e}
\usepackage{float}
\usepackage[utf8]{inputenc}

\usepackage[normalem]{ulem}

\usepackage{siunitx}
\usepackage{tabularx}

\begin{document}
\begin{titlepage}

\title{AI Reasoning Models for Problem Solving in Physics}

\author{Amir Bralin}
\affiliation{Department of Physics and Astronomy,  Purdue University.}

% \author{Amogh Sirnoorkar}
% \affiliation{Center for Advancing the Teaching and Learning of STEM (CATALYST), Purdue University.}

\author{N. Sanjay Rebello}
\affiliation{Department of Curriculum and Instruction, and Department of Physics and Astronomy, Purdue University.}


\begin{abstract}
Reasoning models are the new generation of Large Language Models (LLMs) capable of complex problem solving.
Their reliability in solving introductory physics problems was tested by evaluating a sample of \(n = 5\) solutions generated by one such model---OpenAI's \texttt{o3-mini}---per each problem from 20 chapters of a standard undergraduate textbook.
In total, \(N = 408\) problems were given to the model and \(N \times n = 2,040\) generated solutions examined.
The model successfully solved 94\% of the problems posed, excelling at the beginning topics in mechanics but struggling with the later ones such as waves and thermodynamics.

\clearpage
\end{abstract}

\maketitle
\end{titlepage}

\section{Introduction}
\label{sec:intro}
Traditional education in physics emphasizes solving well-structured, mathematical word problems also called \textit{story problems}~\cite{Jonassen2003}, since they are commonly embedded within a shallow story context.
At the undergraduate level, story problems in physics start as some standard textbook, end-of-chapter problems and exercises.
At the graduate level, they continue playing an important role in student learning, typically as a part of problem sets assigned by the instructors.
There are well-known issues with this kind of problem solving:
``When solving these problems, students are not motivated to search for underlying concepts, but rather are encouraged to look locally for formulas and worked-out examples and then do plug-and-chug to get a correct answer~\cite{Ding2011}.''
% Thus, story problems must be taught with an emphasis on their conceptual structure as opposed to some situational features, which may render students' plug-and-chug activity ineffective.
These issues get exacerbated as technology advances and offers new ways to look up a given problem's solution.
In this paper, we assume that using powerful technology for the sake of learning how to solve story problems in physics is unavoidable, and we explore how it can be done safely and reliably.

Artificial Intelligence (AI) systems are becoming increasingly sophisticated, yet they still exhibit limitations in their reasoning, numerical accuracy, and evaluative capacity---the issues so crucial for problem solving in physics.
One study explored the problem-solving capability of popular AI-based software ChatGPT by OpenAI~\cite{Wang2024}.
In total, 40 real-world physics problems were given to its user interface and the resulting output evaluated.
Of all problems, 24 were under-specified in terms of numerical values given in the problem statement.
ChatGPT failed to make reasonable assumptions about the physical situation described in those problems and estimate the appropriate quantities.
That is, only 2 out of 24 under-specified problems were solved.
For the rest of the problems which were specified by the authors, in contrast, ChatGPT solved 10 (out of 16).
In addition, in both categories of the problems used, ChatGPT committed many calculation mistakes.
This shows the key limitation of the underlying \textit{language model} which ChatGPT at the time was based on: 
it is designed for handling words and sentences, not numbers and formulae.
% Whether it must generate a numerical value on its own (as in estimating a physical quantity) or simply process a given value (as in calculations), this remains a challenge.

Another study also investigated the problem-solving capability of ChatGPT but on a smaller scale~\cite{Kieser2024}.
A single mechanics problem about a point mass sliding down a circular track was given to the user interface as input.
Ten runs were conducted to obtain output along with its statistical variation.
As a result, ChatGPT solved the problem correctly only 5 out of 10 times.
To study its output more systematically, the authors identified four phases of problem solving:
(1) problem representation,
(2) strategy selection,
(3) execution, and
(4) evaluation.
They found that, by default, ChatGPT did not engage in the last phase---evaluating the generated solution to meet the problem's original goal---albeit successfully engaging in the other three.
This could be improved, they noted, by adding a verbal instruction called \textit{user prompt} to the input: ``In particular, describe whether your solution is plausible and and for what reasons you chose your solution.''
% In the domain of AI research and development, constructing such prompts in order to prolong the language model's output in the desired direction was coined ``chain of thought.''

\subsection{Background}
% Language modeling is a computational description of the letters, words, and sentences that occur in \textit{natural languages}--not formal languages such as programming languages and logic systems.
% Hence the term Natural Language Processing (NLP), which is a subfield of AI research and development.
% A given word is likely to occur in a sentence under the specific context.
% Estimating this likelihood with high accuracy is the goal of all language models.
The language models powering such software as ChatGPT are called Large Language Models (LLMs).
They synthesize vast Internet data to produce intelligible linguistic output.
The more data is used in this process and the more computational effort is exerted, the more ``intelligent'' this technology becomes.
Moreover, crossing some threshold in processing so much information may suddenly unlock certain model capabilities such as translating languages or computer coding, akin to phase transitions~\cite{Huberman1987}.
As a result, the AI industry appears to make progress unexpectedly for the general public.

% The Stanford Institute for Human-Centered AI (\href{https://hai.stanford.edu/}{HAI}) reports~\cite{HAI2025}: 
% \begin{quote}
%     Even though the addition of mechanisms such as chain-of-thought reasoning has significantly improved the performance of LLMs, these systems still cannot reliably solve problems for which provably correct solutions can be found using logical reasoning, such as arithmetic and planning, especially on instances larger than those they were trained on.
%     This has a significant impact on the trustworthiness of these systems and their suitability in high-risk applications.''
%     (p. 15)
% \end{quote}
% Taking the fact that education is among the highest-risk human endeavors as granted, it is the goal of this study to contribute toward establishing such \textit{trustworthiness} and suitability for AI problem solving in physics.

LLMs, as artificial neural networks~\cite{Bengio2021}, simply generate the most likely strings of characters in a text, following a given input.
Stable sequences of characters such as prefixes and suffixes, and even some short words, alongside standard characters such as the alphabet letters, numerals, and punctuation symbols, which occur most often in English, are treated as the units of LLM input/output called \textit{tokens}.
All unique tokens stored in a model's memory comprise its ``vocabulary.''
For the state-of-the-art models, this vocabulary size may be quite large: OpenAI's GPT-3.5, which ChatGPT was based on when it was first released on Nov. 30, 2022~\cite{openai2022}, had reportedly around 50k tokens in total.
Later versions have increased this number to 100k and beyond~\cite{arxiv-Yang2024}.
Advanced LLMs are tested against certain benchmarks commonly accepted in the AI research community.
For example, FrontierMath~\cite{arxiv-Math} sets the standard for mathematical capability and GPQA (Google-Proof Q\&A)~\cite{arxiv-GPQA} evaluates the models on physics, chemistry, and biology.
% They have their own issues to consider; see, for example, a review~\cite{arxiv-Eriksson2025}
LLMs loosely referred to as \textit{reasoning models}~\cite{openai2025a} achieve impressive scores on these benchmarks, resulting in a computer-generated output indistinguishable from human text.

% When it comes to problem solving in physics, the likelihood of generating an appropriate sequence of tokens must decrease with difficulty.
% That is, it is trivial for the LLM to find a text-based solution to an easy problem.
% But for harder problems, the sequence of symbols necessary for arriving at a correct solution becomes ever more unlikely to generate.
% The ability to arrive (sometimes, guess) the right solution for a very hard problem, which involves ``self-regulated psychological processes and activities necessary in dynamic environments to achieve ill-defined goals that cannot be reached by routine actions'' and requires ``creative combinations of knowledge and a broad set of strategies''~\cite{DornerFunke2017} remains exclusively in the domain of human experts (see the report~\cite{HAI2025} for more insight).

% The immediate impact of using LLMs, in general and in classroom specifically, is their cost.
% OpenAI charges a certain amount of dollars per token.
% Other, open-source models still need to be used with the help of large computational facilities (such as remote servers).
% They may charge fees too.
% Beside the financial cost, the LLMs require a significant amount of (electrical) energy.
% One can estimate the number of Watts per hour needed for generating a typical problem solution text.
% All of these factors make it paramount that LLM usage is kept under strict supervision.
% Predicting the response time and size for a given prompt may help navigating this costly usage of AI.

\subsection{Motivation}

Knowing what the advanced AI models are capable of will inform educators about their potential harm as well as benefit in using them for teaching and learning physics.
In problem solving, with every progressive step that AI takes, it must be reliable to a high degree in order to be used in the classroom.
The goal of this study was to evaluate the reasoning models in the context of introductory physics.
Our working definition of \textit{reliability} was the ability of a model to solve correctly and repeatedly story problems on a given physics topic.
The model performance by individual topic provides a more precise view of the relevant model properties when compared with typical single-number benchmarks used in the broader AI community.
Our guiding questions were:
\begin{enumerate}
    \item How reliable are AI reasoning models when solving story problems in physics?
    \item What is the distribution of the story problem-solving ability of AI reasoning models across standard topics in physics?
\end{enumerate}
As presented in the following sections, the state-of-the-art AI reasoning models may be reliable problem-solvers in the beginning topics of a typical introductory physics course, but still struggle with solving problems in later topics such as waves and thermodynamics.
% This may be due to them being of higher difficulty which involve abstract thinking and precise calculations.
% Or it may be simply because there is more data present in training theses models about mechanics problems when compared with more advanced topics such as heat and electromagnetism.

\section{Methods}

Among various available Large Language Models (LLMs), the family of reasoning models called the ``o-series'' by OpenAI was selected for this study due to the popularity of ChatGPT.
This family started with a preview of its first reasoning model \texttt{o1}~\cite{openai2024}.
Soon, it included an updated version \texttt{o3} working at full capacity, as well as a ``lightweight'' version \texttt{o3-mini}.
Specifically, the model \texttt{o3-mini}~\cite{openai2025b} was used in this study due to its \textit{affordability} and its availability at the time of data collection and analysis.
% Note that yet another updated reasoning model \texttt{o4-mini} was released on Apr 16, 2025.
The access was provided by OpenAI's Application Programming Interface (API) at \url{https://openai.com/api/}.
The cost of using \texttt{o3-mini} was listed as \$1.1 per million tokens for processing input and \$4.4 per million tokens for generating output.
We estimated the total number of input tokens, which represent all problems in our dataset, to be no more than 1M tokens.
The total number of output tokens, which represent the solutions generated by the model, was around 3.3M tokens.
Thus, the total cost was projected to be around \$15.
The actual cost was greater because some errors during this entire process were unavoidable and we had to rerun the model after fixing each issue encountered.

Among various sources of physics problems found in standard textbooks, ``Fundamentals of Physics'' Vol. 1 by Halliday and Resnick~\cite{HR2022} was selected due to its status and popularity in the undergraduate curriculum.
The table of contents is shown in Table~\ref{tab:results}.
There are 20 chapters in total, spanning the standard topics in mechanics (including the kinetic theory) and classical thermodynamics.
Column ``Odd-numbered Problems'' lists the total number of all problems for which the answer is given in each chapter.
Column ``Text-only Problems'' lists the numbers of text-based problems only, which were used for analysis.
The bottom row shows the sums of all problem numbers above it for each category.
Column ``Problems Solved'' is related to the results of this study described in the next section.

Odd-numbered problems from each chapter, for which the answer key is available, were copied into {\LaTeX} and then into a Python code, which simply sent the problem text to OpenAI API and received the resulting solution text repeatedly.
% Some problems in the textbook included figures and tables.
% Since \texttt{o3-mini} only handles text, they were all ignored.
Since \texttt{o3-mini} only handles text, all problems that contain a figure or a table were ignored, even if the information was also given in the problem text.
Thus, out of the total 629 (odd-numbered) problems from all 20 chapters, only \(N = 408\) text-based problems were selected and used for this study.
% Note that the textbook answers---whether for text-based problems or those with a figure/table---are given in text: a numerical value, a word or two, or a combination of both.
% Thus, to expand our analysis beyond the text-based problem solving (see Sec.\ref{sec:limits}) one will only need to have a reasoning model able to process images as input.
% The output may remain text-only.

The reasoning model \texttt{o3-mini} was given each problem statement as input without any additional instructions.
It produced each solution as its output, which then was evaluated for \textit{correctness} by comparing the final answer in the solution to the textbook's answer key.
This was done \textit{manually}, by a human expert.
The textbook answers themselves were not evaluated for correctness.
We assume that a book with so many editions has identified any potential errors in its answer key.
% Thus, the model's validity was established within the scope of this study.

The prompt given to \texttt{o3-mini} simply contained the problem text copied from the textbook (in the {\LaTeX} format) and nothing else.
In general, problem solving with LLMs requires well-structured prompts.
This is no longer required for reasoning models however.
As OpenAI suggests, the latter ``provide better results on tasks with only high-level guidance'' and the former ``benefit from very precise instructions~\cite{openai2025c}.''

To establish the model's reliability, the solution output was generated \(n_{\text{sample}} = 5\) times for each problem.
In order for a given problem to be counted as ``successfully solved'' by the model, all generated solutions must have resulted in the correct answer, according to the textbook's key.
Those problems for which the generated solutions were either completely or partially correct were marked as ``not solved'' and saved for further analysis: the solution text was examined in order to identify some common properties resulting in failure.

% In addition, the time or ``effort'' required by the LLM to generate a complete solution to a problem was estimated by the number of tokens produced as a result.
% We assume that the model takes the same amount of time and computational power to produce any token from its vocabulary.
% The number of ``reasoning'' tokens, which remains \textit{hidden} under the API, provides an estimate of ``internal'' effort spent by the model before arriving at an appropriate solution.
% The total number of \textit{visible} characters involved in the model response provides an estimate of the final solution size.
% Given a problem, it may involve several steps (such as (a) to (d)) that will make its solution artificially long;
% the effective solution size must be the total size divided by this number.

\begin{table*}[tbhp]
    \caption{Textbook chapter titles and the corresponding numbers of problems.}
    \centering
    % \begin{ruledtabular}
    \begin{tabular}{ l  c  c  c }
        % \hline
        \textbf{Chapter} & \textbf{Odd-numbered Problems} & \textbf{Text-only Problems} & \textbf{Problems Solved} \\
        \hline
        \hline
        \phantom{0}1. Measurement & 16 & 13 & 13 (100\%) \\
        \phantom{0}2. Motion Along a Straight Line & 35 & 25 & 25 (100\%) \\
        \phantom{0}3. Vectors & 22 & 17 & 17 (100\%) \\
        \phantom{0}4. Motion in Two and Three Dimensions & 41 & 32 & 30 \phantom{0}(94\%) \\
        \phantom{0}5. Force and Motion--I & 34 & 18 & 16 \phantom{0}(89\%) \\
        \hline
        \phantom{0}6. Force and Motion--II & 30 & 15 & 15 (100\%) \\
        \phantom{0}7. Kinetic Energy and Work & 26 & 17 & 17 (100\%) \\
        \phantom{0}8. Potential Energy and Conservation of Energy & 33 & 12 & 11 \phantom{0}(92\%) \\
        \phantom{0}9. Center of Mass and Linear Momentum & 40 & 22 & 22 (100\%) \\
        10. Rotation & 34 & 25 & 24 \phantom{0}(96\%) \\
        \hline
        11. Rolling, Torque, and Angular Momentum & 35 & 18 & 18 (100\%) \\
        12. Equilibrium and Elasticity & 26 & \phantom{0}7 & \phantom{0}6 \phantom{0}(86\%) \\
        13. Gravitation & 35 & 26 & 25 \phantom{0}(96\%) \\
        14. Fluids & 36 & 25 & 25 (100\%) \\
        15. Oscillations & 32 & 20 & 19 \phantom{0}(95\%) \\
        \hline
        16. Waves--I & 30 & 23 & 20 \phantom{0}(87\%) \\
        17. Waves--II & 35 & 29 & 22 \phantom{0}(76\%) \\
        18. Temperature, Heat, and the First Law of Thermodynamics & 33 & 23 & 22 \phantom{0}(96\%) \\
        19. The Kinetic Theory of Gases & 32 & 25 & 23 \phantom{0}(92\%) \\
        20. Entropy and the Second Law of Thermodynamics & 24 & 16 & 14 \phantom{0}(88\%) \\
        \hline
        & \textbf{629} & \textbf{408} & \textbf{384} (\textbf{94\%})\\
        % \hline
    \end{tabular}
    % \end{ruledtabular}
    \label{tab:results}
\end{table*}

\section{Results}
% The table of contents from the textbook is shown on Table~\ref{tab:results}.
% Column ``Odd-numbered Problems'' lists the total number of all problems for which the answer is given in each chapter.
% Column ``Text-only Problems'' lists the numbers of text-based problems only, which were used for analysis.
% The bottom row shows the sums of all problem numbers above it for each category.
In Table~\ref{tab:results}, column ``Problems Solved'' shows the percentage of (text-only) problems the AI reasoning model \texttt{o3-mini} successfully solved for each chapter.
It drops visibly for the last portion of the topics: chapters 15--20.
Especially, the topic of Waves posed a challenge to the model.
Chapter 16 had 87\% of its problems solved (20 out of 23), and Chapter 17 only 76\% (22 out of 29).
Though the chapter on Equilibrium and Elasticity also had a lower percentage, its sample size was much smaller and thus not comparable.
% Chapter 5 and 20 also resulted in somewhat lower proportions of solved problems (having 18 and 16 problems given to the model, respectively).

In total, there were 24 problems that were not successfully solved (that is, about 6\%) by \texttt{o3-mini} in our setup.
Examples below show the typical errors that led to such failure.

% \begin{quote}
%     \textbf{Ch. 4, Problem 55}:
%     A ball rolls horizontally off the top of a stairway with a speed of \qty{1.52}{m/s}.
%     The steps are \qty{20.3}{cm} high and \qty{20.3}{cm} wide.
%     Which step does the ball hit first?
% \end{quote}

\begin{quote}
    \textbf{Ch. 4, Problem 63}:
    At $t_1 = 2.00$ \unit{s}, the acceleration of a particle in counterclockwise circular motion is $6.00 \hat{\textbf{i}} + 4.00 \hat{\textbf{j}}$ \unit{m/s^2}.
    It moves at constant speed.
    At time $t_2 = 5.00$ \unit{s}, the particle's acceleration is $4.00 \hat{\textbf{i}} - 6.00 \hat{\textbf{j}}$ \unit{m/s^2}.
    What is the radius of the path taken by the particle if $t_2 - t_1$ is less than one period?
\end{quote}

To solve this problem, one must consider several possible options.
The dot product between the two given vector values of acceleration results in 0, indicating that the particle has moved by $\pi/2 + k\pi$ (where $k=0, 1, 2, \dots$) \unit{rad} on the circle.
The problem states that the time passed is less than a period, which means that it moved by either $\pi/2$ or $3\pi/2$ \unit{rad}.
The model \texttt{o3-mini} considered only the first option and arrived at the wrong answer.

\begin{quote}
    \textbf{Ch. 13, Problem 41}:
    Two neutron stars are separated by a distance of \qty{1.0e10}{m}.
    They each have a mass of \qty{1.0e30}{kg} and a radius of \qty{1.0e5}{m}.
    They are initially at rest with respect to each other.
    As measured from that rest frame, how fast are they moving when (a) their separation has decreased to one-half its initial value and (b) they are about to collide?
\end{quote}

This is an example of typical language model behavior.
% The problem involves multiplication with power operations.
Here, \texttt{o3-mini} simply committed a calculation error: ``Taking the square root: $v = \sqrt{3.335 \times 10^{14}} \approx 5.78 \times 10^5$ \unit{m/s}.''
The correct result should be $1.826 \times 10^7$ \unit{m/s}.

\begin{quote}
    \textbf{Ch. 19, Problem 11}:
    Air that initially occupies \qty{0.140}{m^3} at a gauge pressure of \qty{103.0}{kPa} is expanded isothermally to a pressure of \qty{101.3}{kPa} and then cooled at constant pressure until it reaches its initial volume.
    Compute the work done by the air.
    (Gauge pressure is the difference between the actual pressure and atmospheric pressure.)
\end{quote}

This involves isothermal expansion and logarithmic operation that the model worked out without any issues.
The inconsistency arose at the last run when it suddenly ended up with an answer (\qty{1.8}{J}) different from the previous four attempts (\qty{5.6}{kJ}, as listed in the answer key).
Upon closer examination of the solution output, the issue was the key assumption of the problem: that the second value of pressure given in the problem statement is the actual pressure in the gas and not the gauge value.
This way, the gas isothermally expanded from a pressure of $P_1 = P_\text{atm} + P_\text{gauge} = 101.3 + 103.0 = \qty{203.3}{kPa}$ to a pressure of $P_2 = \qty{101.3}{kPa}$.
The logarithmic relation $\ln{P_1/P_2}$ yields the ratio between the corresponding volumes of the gas $V_2/V_1$, necessary for the consequent solution steps that result in the correct answer.
The model in its first four runs made this assumption and successfully solved the problem.
The output read: ``Since no `gauge' is mentioned for this step, we interpret 101.3 kPa to be the absolute pressure.''
In the last attempt, however, it made did not make this assumption and took the value $P_2$ to be another gauge value.
We counted the model solution to be overall unsuccessful due to the inconsistency in its responses.
It must be noted, however, that this type of errors may be due to the problem statement rather than the model's problem-solving capability.

% \begin{table}[tbh]
%     \caption{Problems solved by \texttt{o3-mini} in each chapter.}
%     \centering
%     % \begin{ruledtabular}
%     \begin{tabularx}{\columnwidth}{ X | c  }
%         \textbf{Chapter} & \textbf{Problems Solved (\%)} \\
%         \hline
%         \phantom{0}2. Motion Along a Straight Line & 100 \\
%         \phantom{0}4. Motion in Two and Three Dimensions & 94 \\
%         \hline
%         \phantom{0}7. Kinetic Energy and Work & 100 \\
%         \phantom{0}8. Potential Energy and Conservation of Energy & 92 \\
%         \hline
%         \phantom{0}5. Force and Motion--I & 89 \\
%         \phantom{0}6. Force and Motion--II & 100 \\
%         \hline
%         15. Oscillations & 95 \\
%         16. Waves--I & 87 \\
%         17. Waves--II & 76 \\
%         \hline
%         18. Temperature, Heat, and the First Law of Thermodynamics & 96 \\
%         19. The Kinetic Theory of Gases & 92 \\
%         20. Entropy and the Second Law of Thermodynamics & 88 \\
%     \end{tabularx}
%     % \end{ruledtabular}
%     \label{tab:distribution}
% \end{table}

\section{Discussion and Conclusions}

Despite significant improvement, reasoning models are, by their design, statistical language models.
They require a lot of data to be trained on, using sophisticated machine learning algorithms.
For the kind of data that was the subject of this study---story problems from introductory physics---there may be enough of it in the training of these models.
Then the high accuracy in solving them is expected.
The instances of failure to solve particular problems in this dataset then must be due to some inherent features of the reasoning model used---OpenAI's \texttt{o3-mini}.
% Upon a closer examination of the Large Language Model (LLM) design, these features seem to be present in all models following the common architecture.
They fall into one of the two broad categories of errors:
(1) the model simply follows the verbal reasoning it generates, without any way to evaluate its intermediate steps by other means (such as physics simulation),
(2) the model commits simple calculation and rounding mistakes without any way to evaluate its numerical results by other means (such as math computation).
Both of these error sources may be fixed by adding specific tools for simulation and computation to the model's workflow, a process called \textit{augmenting} the model.
While there has been some progress in this direction, a thorough and comprehensive initiative is yet to be undertaken (see Ref.~\cite{Mialon2023} for a survey by Meta AI).
Until then, the question of reasoning models' reliability for the purposes of problem solving will remain open.

% In this study with a particular model \texttt{o3-mini}, it solves the great majority of the given problems consistently.
The particular reasoning model \texttt{o3-mini} consistently solved the great majority of the problems given in this study.
Perhaps it could be used for solving relatively easy problems that some students, nevertheless, find challenging.
This utility must be in the context of learning from \textit{worked examples}.
Students should use this powerful tool of reasoning models only when they need to see some worked examples in order to learn how to solve the problem for which the examples are generated.
Without such a diligent, conscientious approach to reasoning models (and all LLMs in general), they quickly degrade to being used as a tool for cheating, plagiarism, and misinformation.

% Whether using story problems or their ill-structured counterparts for teaching physics, both types require a rich and diverse pool of \textit{worked examples} to be used by learners.
According to extensive findings from research on this so-called worked-example effect, reviewed in article~\cite{Atkinson2000}, students really do benefit from detailed solutions to problems presented to them in the process of learning a given topic.
However, as the authors point out, there are certain aspects of worked examples that demand careful consideration:
First, they must be designed and presented so that the students' problem solving improves with time and effort;
several (at least, two) examples per problem should be designed to elucidate the problem's complex structure;
each problem as well as every example related to it should be presented in a clear, unified format (integrating visuals, sound, and text if applicable) as to minimize cognitive overload~\cite{Sweller1988}.
Second, the examples relevant to a given problem must be properly coordinated so as to enhance student learning;
say, each pair of examples connected to a given problem must be studied as a separate block before proceeding to the next problem with its own pair of examples.
% ``... lessons that pair each worked example with a practice problem and intersperse examples throughout practice will produce better outcomes than lessons in which a blocked series of examples is followed by a blocked series of practice problems.'' (p. 195,~\cite{Atkinson2000})
Third, the students themselves vary in their ability to learn from examples, so this too must be addressed in a given lesson;
notably, self-explanations~\cite{Chi1989}, which are key to successful problem solving, may be directly taught to students and promoted in the class.

Whether the problem solutions provided by LLMs exhibit the mentioned properties of worked examples depends on their design.
They are not natural phenomena but rather powerful tools built with a specific ``architecture'' and for a specific purpose.
% Designed one way, they may impede student learning by giving away the right answer too soon or by showing wrong technique in solving a problem; designed another way, they may benefit education...
The vast Internet data used for training these models serves as the base.
Then, AI companies recruit experts from various fields of study (they are referred to as ``human annotators'') to chat with a given model and provide it with some template to follow.
This stage may vary in the specific techniques that AI developers use for calibrating the model, but the main idea is that people guide the model behavior, one way or another~\cite{arxiv-Ouyang2022}.
Therefore, it is conceivable that future advanced AI models will be developed based on design principles from STEM education research.

The distribution of a reasoning model's problem-solving ability across the standard physics topics was also explored in this study.
Overall, it may seem quite uniform, with the percentage of successfully solved problems averaging around 94\%.
When examined closer, however, a subtle pattern was observed: the model performs worse on later chapters than on the earlier ones.
% 100\% for Ch. 2 ``Motion in One Dimension,'' while 94\% for Ch. 4 ``Motion in Two and Three Dimensions.''
% 100\% for Ch. 7. ``Kinetic Energy and Work,'' while 92\% for Ch. 8. ``Potential Energy and Conservation of Energy.''
% The outlier is the pair of Ch.'s 5 and 6 (``Force and Motion--I'' and ``II''), which displays the opposite relation (the accuracy rises from 89\% to 100\%).
% This is worth investigating in more detail.
The performance drops for chapters 16 and 17 on waves since they involve more detailed calculations, due to their underlying mathematical apparatus.
% Similarly, we observe that the model performance is gets worse for related chapters:
% 95\% for Ch. 15. ``Oscillations,'' 87\% for Ch. 16 ``Waves--I,'' and 76\% for Ch. 17 ``Waves--II.''
% Finally, the results drop from 96\% to 92\% to 88\% fro the Ch.'s 18. ``Temperature, Heat, and the First Law of Thermodynamics,'' 19. ``The Kinetic Theory of Gases,'' and 20. ``Entropy and the Second Law of Thermodynamics,'' respectively.
Then, as the topics transition to kinetic theory and thermodynamics in chapters 18--20, the model performs gradually worse again (dropping from 96\% to 92\% to 88\%, respectively).
This might be due to the increasing complexity of each consecutive topic.
Alternatively, these topics may be underrepresented on the Internet (compared with mechanics) and thus contribute less to the AI model training.

As these models are updated and reach ever higher performance on this and similar tests of problem-solving accuracy and reliability, the exact threshold to demand from them will be a matter of convention.
We may demand that an AI model has to solve all 100\% (or some other floating-point number with a very small margin of error) of the problems that it might encounter in a standard physics course before it is deployed in that course, whether for assessment or tutoring.

% In principle, we hold that AI use in classroom must be strictly regulated.
% For example, when a teacher decides to let the students use a reasoning model for their homework problems (say, in physics), it is the teacher's duty to verify that the model is able to solve each problem reliably beforehand.
% We are aware of the grand challenges to this approach.
% For one, AI technology is widely available outside the classroom today.
% Meeting this challenge will require governmental control and a broader social support for the corresponding measures, which goes beyond the scope of this work.

\section{Limitations and Future Work}
\label{sec:limits}

A natural continuation of this study is to expand our analysis to further problems from ``Fundamentals of Physics'' Vol. 2 that includes Electromagnetism, Optics, Relativity, and Modern Physics.
This way, we will have a broader view of the AI capability and reliability in physics problem solving.
Increasing the depth of this view will demand other, more challenging story problems to be considered.
For example, advanced undergraduate or graduate-level problems may be used.
The question of why the model performance of \texttt{o3-mini} was dropping as the topics progressed remains open.
% To answer it, we will need to...
A new, updated, and more capable reasoning model \texttt{o4-mini} was released by OpenAI on Apr. 16, 2025~\cite{openai2025d}.
We expect it to achieve even higher accuracy on solving the problems considered in this study.
% However, unless its underlying structure of language modeling was changed in order to account for handling arbitrary numbers, which are not in the model's vocabulary, this model will still produce unreliable output when calculating numerical values for its final answers and intermediate values.
In addition, this new model is able to process images alongside text.
% This means that virtually all odd-numbered problems from the textbook may be tested.
This entire endeavor, nevertheless, is limited by the type of problems so popular in physics: story problems.
Evaluating the properties of AI models in solving other problem types is also much needed if they are to be used within such contexts.

\section*{Acknowledgment}

We thank Razan (Rosie) Hamed and Syed Furqan Hashmi for their feedback on the manuscript draft.
This work was supported by U.S. National Science Foundation (NSF) Grant 2300645.
All opinions, results, and findings expressed here are those of the authors and not of the NSF. 

\clearpage
\bibliography{source}

\end{document}